# Nonreciprocal low-noise acoustoelectric microwave amplifiers with net gain in continuous operation

Lisa Hackett[1], Michael Miller[1], Scott Weatherred[1], Shawn Arterburn[1], Matthew J. Storey[1], Greg Peake[1], Daniel Dominguez[1], Patrick Finnegan[1], Thomas A. Friedmann[1], and Matt Eichenfield[1]

1. Microsystems Engineering, Science, and Applications, Sandia National Laboratories, Albuquerque, NM 87123, USA
Correspondence: Matt Eichenfield (meichen@sandia.gov)

**Abstract**
Over sixty years ago, it was hypothesized that specially designed acoustic systems that leveraged the acoustoelectric effect between phonons and charge carriers could revolutionize radio frequency electronic systems by allowing nonlinear and nonreciprocal functionalities such as gain and isolation to be achieved in the acoustic domain. Despite six decades of work, no acoustoelectric amplifier has been produced that can achieve a large net (terminal) gain at microwave frequencies with low power consumption and noise figure. Here we demonstrate a novel three-layer acoustoelectric heterostructure that enables the first-ever continuously operating acoustoelectric amplifier with terminal gain at gigahertz frequencies. We achieve a terminal gain of 11.25 dB in a 500 µm long device, operating at 1 GHz with a DC power dissipation of 19.6 mW. We also realize broadband gain from 0.25-3.4 GHz and nonreciprocal transmission exceeding 44 dB at 1 GHz. Our acoustic noise figure is 2.8 dB, which is the lowest-ever demonstrated noise figure for an acoustoelectric amplifier. We discuss generally how to optimize these acoustoelectric heterostructures and show that it should be immediately achievable to produce devices with even larger gain in shorter lengths while simultaneously having lower power consumption and noise figure.

It has been known since the 1960's that acoustoelectric interactions of propagating phonons and electrons can lead to nonreciprocal amplification of radio frequency (RF) acoustic waves,[1-4] potentially allowing the same piezoelectric acoustic wave devices that are used as filters in radios to also become the amplifiers, as well as produce devices such as isolators and circulators. Given the small feature sizes associated with acoustic wavelength-scale structures at microwave frequencies, as well as the already ubiquitous use of acoustic RF filters, the prospect of combining the filtering, amplifying, and nonreciprocal functions into one and the same device offers a degree of miniaturization that is unprecedented, especially when compared to the current trend of co-packaging piezoelectric acoustic filters with semiconductor-based amplifiers and gyromagnetic nonreciprocal devices. Despite 60 years of work and technology maturation, demonstrating a nonreciprocal acoustoelectric amplifier that operates continuously at gigahertz frequencies while producing net (terminal) gain—i.e., more gain than insertion loss—has remained elusive. This deficit has been generally due to a host of issues, such as weak interactions between the piezoelectric acoustic wave and semiconductor carriers,[5-7] poor thermal conductivity causing deletrious heating effects before large gain can be achieved,[8-12] and the difficulty of integrating high-quality semiconductor materials with strongly piezoelectric materials such as lithium niobate ($LiNbO_3$).[13-15]

Here, for the first time, we demonstrate a nonreciprocal acoustoelectric amplifier operating continuously at 1 GHz, while producing a net gain of 11.25 dB in a 500 μm long amplifying delay line and a nonreciprocal transmission of 44 dB. Moreover, the gain bandwidth exceeds 3 GHz, and the acoustic noise figure at 1 GHz is 2.8 dB, which is the lowest-ever demonstrated noise figure for an acoustoelectric amplifier. Thus, the device constitutes the first technologically viable acoustoelectric amplifier and the largest continuously generated surface acoustic wave nonreciprocity reported by any method. The amplifier is able to achieve this long-sought performance using a sophisticated acoustoelectric heterostructure consisting of 1) an ultra-thin compound semiconductor (indium gallium arsenide ($In_{0.53}Ga_{0.47}As$)) with low electrical conductivity ($\sigma$) and high-mobility ($\mu$), 2) a thin, strongly piezoelectric film ($LiNbO_3$) directly underneath the semiconductor, and 3) a substrate (silicon) that enhances the confinement of the phonons in the thin piezoelectric film to increase electromechanical copuling, provides a high thermal conductance medium for the removal of dissipated heat, and has low dielectric RF loss.

In this article, we present our acoustoelectric heterostructure, experimentally characterize high-performance amplifiers built in that heterostructure architecture, and show generally how to optimize the system for active acoustic wave devices. The potential to further increase gain while decreasing the amplifier size, power consumption, and noise figure is discussed. When combined with recent demonstrations of acoustoelectric circulators and switches,[10,16-18] these devices pave the way towards all-acoustic RF signal processing front-ends that can be made on a single chip, which could lead to both higher performance and significant miniaturization of RF systems. The potential reduction in size, weight, and power has the potential to transform wireless broadband communcation, Wi-Fi, Bluetooth, GPS, RF identification, and wireless electrical biosensors. Beyond wireless RF devices, the heterostructure approach we present here for acoustoelectric devices, with the resulting experimental advances, could have an impact in fields such as non-reciprocal electronics,[19-22] optomechanics[23-25] and quantum information processing[26-31] where piezoelectric acoustic wave technology is increasingly utilized for advanced and novel functionalities. As discussed above, the acoustoelectric interaction in these devices is inherently nonreciprocal for acoustic waves propagating parallel and antiparallel to the drifting charge carriers in the semiconductor. In this work, we show that exceptional levels of nonreciprocal transmission can be achieved using the acoustoelectric effect with surface acoustic waves, fully 2500 times larger than the previous state-of-the-art using nonlinear parity-time symmetric resonators,[22] while simultaneously being broadband and requiring no external integrated circuits. Building upon the nonreciprocal directional amplifier presented here, this work could ultimately lead to novel approaches for non-Hermetian phononics, such as acoustic wave isolation and routing, in compact and continuously operating devices at microwave frequencies with low noise.

**Acoustoelectric heterostructure for active piezoelectric acoustic wave devices**

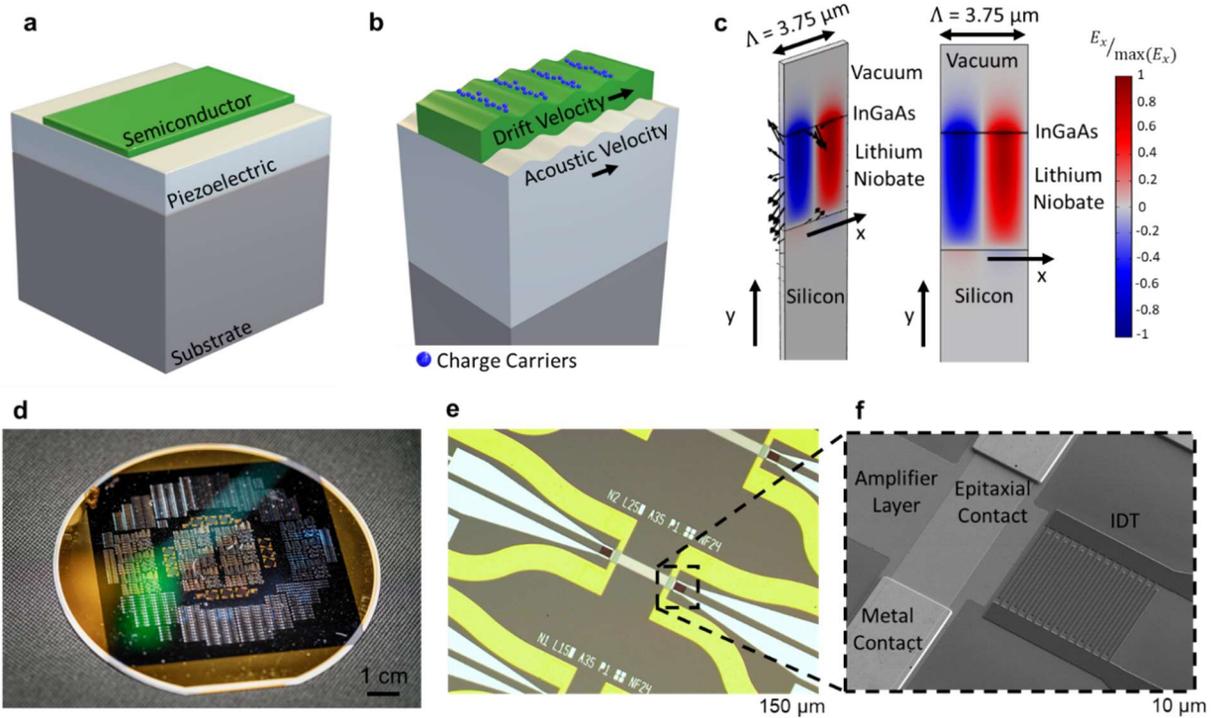

**Fig. 1: Material platform overview.** (a) Schematic of a three-layer heterostructure to support high-performing active acoustic wave devices based on the acoustoelectric effect and (b) schematic of the interaction between the semiconductor charge carriers and piezoelectric acoustic wave. (c) Longitudinal electric field model in our material stack of an $In_{0.53}Ga_{0.47}As$ semiconducting film on a $LiNbO_3$ piezoelectric film on bulk silicon. The black arrows indicate material displacement. The acoustic wavelength ($\Lambda$) is 3.75 μm, the $LiNbO_3$ thickness is 5 μm, and the $In_{0.53}Ga_{0.47}As$ thickness is 50 nm with a space charge density ($\rho$) given by $\rho = qN_d$ where $q$ is the elementary charge and $N_d$ is the semiconductor doping concentration, set to $1 \times 10^{16}$ cm$^{-3}$ for this model. (d) Camera image of the heterostructure wafer after processing. (e) Microscope image of a fabricated acoustic wave amplifier with (f) a scanning electron micrograph to show additional details of the IDT, metal contact, and epitaxial contact.

A schematic of a general three-layer heterostructure that enables high-performance acoustic wave amplification is shown in Fig. 1(a). In this work we focus on a specific implementation with an $In_{0.53}Ga_{0.47}As$ semiconducting layer, a $LiNbO_3$ piezoelectric film, and a silicon substrate.

An illustration of the acoustoelectric effect, which occurs when semiconductor charge carriers interact with a piezoelectric acoustic wave, is shown in Fig. 1(b). In the heterostructure reported here, the interaction occurs between the evanescent longitudinal electric field of the piezoelectric acoustic wave that penetrates the semiconducting film and the charge carriers therein. A finite element method (FEM) model of the longitudinal electric field of the guided piezoelectric acoustic wave with primarily shear-horizontal (SH) acoustic polarization in the $In_{0.53}Ga_{0.47}As$-$LiNbO_3$-silicon heterostructure is shown in Fig. 1(c). The evanescent overlap between the longitudinal electric field in the $LiNbO_3$ and the 50 nm thick $In_{0.53}Ga_{0.47}As$ semiconductor layer enables the acoustoelectric effect in this material stack.

An image of the fabricated heterostructure wafer with acoustic wave amplifiers and delay lines is shown in Fig. 1(d). These devices are enabled by a microfabrication process that is described in detail in the Methods. Wafer bonding provides intimate contact between the $LiNbO_3$ and $In_{0.53}Ga_{0.47}As$. The interdigital transducer (IDT) launches and receives acoustic waves, while quasi-Ohmic electrical contact is made to the $In_{0.53}Ga_{0.47}As$ amplifier layer via a mesa contact structure. The contact mesa comprises a silver/gold electrode layer on an epitaxial contact heterostructure that provides vertically graded doping from the amplifier layer to the metal. A microscope image of an acoustic wave amplifier is shown in Fig. 1(e), and a scanning electron micrograph showing the IDT, mesa contact structure, and amplifier layer is shown in Fig. 1(f).

With the heterostructure approach, we have increased flexibility to tailor the material parameters of each layer, which allows us to realize large gain in a small device footprint while simultaneously achieving low dissipated power and noise figure. We select $In_{0.53}Ga_{0.47}As$ for the semiconductor layer because it has a high bulk $\mu$ of 10,000 $cm^2$/V-s and can be grown by metal-organic chemical vapor deposition (MOCVD) within an $In_{0.53}Ga_{0.47}As$/indium phosphide (InP) lattice-matched stack with controllable thickness and doping and with low defectivity.[32] We show here that, even for a thin layer (~50 nm) in an acoustoelectric heterostructure, the $In_{0.53}Ga_{0.47}As$ $\mu$ can exceed 4000 $cm^2$/V-s. This semiconducting material is also able to bond to $LiNbO_3$ using an ultra-thin (5 nm) InP non-intentionally doped (NID) intermediary layer, which prevents any degradation of device performance on account of being 1/1000$^{th}$ the acoustic wavelength. We select $LiNbO_3$ for the piezoelectric layer because, as we report here, exceptionally large electromechanical coupling coefficients ($k^2$) exceeding 10% and low acoustic loss can be achieved in YX $LiNbO_3$ films on silicon for a specially chosen acoustic mode with primarily SH polarization. Silicon is used as the substrate material due to its >30X improvement in thermal conductivity compared to bulk $LiNbO_3$.[33,34] The higher thermal conductivity greatly contributes to our acoustic wave amplifier's ability to run with a continuously applied drift field while achieving terminal gain. The high acoustic velocity in silicon supports a guided acoustic wave in the $LiNbO_3$ film. In addition, silicon substrates can be made with high resistivity (>10kΩ), which enables low RF dielectric loss that can cause excess loss for piezoelectric acoustic waves propagating in the $LiNbO_3$.

In this section, we have illustrated a general three-layer acoustoelectric heterostructure that can be used to develop high-performance acoustic wave amplifiers, presented our specific $In_{0.53}Ga_{0.47}As$-$LiNbO_3$-silicon material platform, and introduced the most critical material parameters for device operation. In the remainder of the article, we report the acoustic wave amplifier performance in terms of gain and noise figure followed by a more detailed discussion on our approach to heterostructure design and how to achieve even larger gain with smaller footprint, power dissipation, and noise figure.

**Acoustic wave amplifier gain and noise figure**

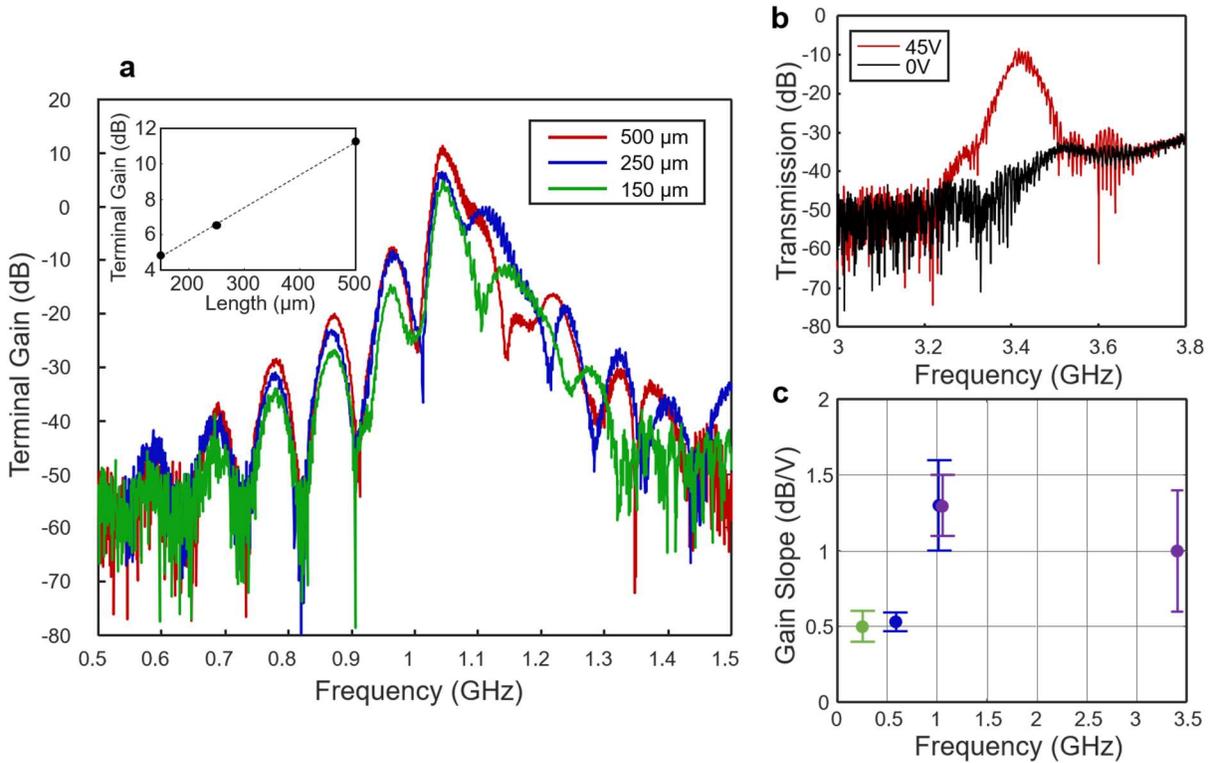

**Fig. 2: Acoustic wave amplifier characterization.** (a) Measured terminal gain as a function of frequency for three different devices with lengths of 150 μm, 250 μm, and 500 μm. For these devices, the acoustic wavelength is 3.75 μm, the IDT aperture is 35 μm, and each IDT has 12 electrode pairs. The inset shows the terminal gain at 1.04 GHz as a function of the device length. For the 500 μm long device, a terminal gain of 11.25 dB is achieved at 1.04 GHz with a dissipated power of 19.6 mW. (b) Measured transmission increase at the third harmonic of an IDT designed to operate at approximately 1 GHz as a function of frequency with and without the application of 45 V. (c) Measured gain slope as a function of frequency. Error bars correspond to ± one standard deviation based on measurements from multiple devices. The three different marker colors correspond to data taken from three different wafers.

A plot of terminal gain as a function of frequency for acoustic wave amplifiers with three different lengths is shown in Fig. 2(a). The technique and experimental setup for the gain measurements are described in the Methods. Terminal gain is achieved only when the gain exceeds the total end-to-end losses across the device, including the losses from the IDTs. The inset of Fig. 2(a) shows the terminal gain at 1.04 GHz as a function of the device length. A terminal gain of 11.25 dB is achieved at 1.04 GHz with a device length of 500 μm and dissipated power of 19.6 mW. In addition, a plot of the transmission as a function of frequency around a 3.4 GHz resonance, with and without an applied bias of 45 V, is shown in Fig. 2(b). This frequency corresponds to the third harmonic of the IDT designed for operation around 1 GHz. While terminal gain is not achieved, there is a transmission increase of 40 dB at 3.4 GHz with the applied bias. These results suggest that terminal gain is achievable at higher frequencies contingent on IDT optimization to minimize excess insertion losses. Figure 2(c) shows the measured gain slope as a function of operating frequency from devices measured across three different wafers. Broadband gain from 0.25 – 3.4 GHz with a gain slope exceeding 0.5 dB/V is achieved in this material platform. As can be seen, these types of amplifiers can achieve gain across an exceptionally large bandwidth, demonstrating the feasibility of ultrabroadband acoustoelectric amplifiers in a heterostructure with fixed film thickness. As described in Supplementary Note 1, we also evaluate the gain compression for an acoustoelectric amplifier operating at 250 MHz by measuring the acoustic output power as a function of the acoustic input power. A gain compression of 1 dB is observed at an acoustic output power of 12 dBm when operating at maximum bias. Additional improvements to thermal management should allow these devices to operate with even higher saturation output power.

The maximum operating frequency for an acoustoelectric amplifier in our heterostructure is determined by charge carrier diffusion effects[35], the minimum feature size with which an IDT can be patterned, and, for the particular guided-wave mode used and modeled in this work, the decreasing of $k^2$ with increasing frequency; however, at very high frequencies Rayleigh and SH-surface acoustic wave (SAW) quasi-surface modes can be utilized and have higher $k^2$ than the guided wave mode in this regime. Also, as described in Supplementary Note 2, with increasing frequency the InP adhesion layer can become an appreciable fraction of the acoustic wavelength (thickness/wavelength > 0.01), thereby reducing the evanescent overlap of the evanescent field with the carriers (reducing the effective $k^2$) and, in turn, reducing the gain. While we generally seek to maximize the achievable gain, the variation of gain with adhesion layer thickness potentially allows a tailorable frequency-dependent gain, such as a flat broadband gain response, given that gain typically increases with frequency.

We next assess the noise figure of our acoustoelectric platform. This was done using an amplifier with the lowest possible metal-semiconductor contact resistance, which gives the most accurate noise figure measurement. A plot of the measured terminal gain as a function of frequency with increasing drift field values for this device is shown in Fig. 3(a). A terminal gain of 6.5 dB is achieved at a drift field of 1 kV/cm. As shown Supplementary Note 3, we also demonstrate a nonreciprocal transmission greater than 44 dB in this device. The semiconductor and electromechanical properties of the heterostructure are: a semiconductor thickness of 58 nm, μ of 4220 cm$^2$/V-s, and $k^2$ of 14% (see Methods). A plot of the acoustic noise figure as a function of electronic gain is shown in Fig. 3(b). Also shown is the corresponding terminal noise figure, including the impact of the IDTs, which are lossy and thus substantially increase the noise figure of the system. The noise figure was measured by turning a calibrated noise source on and off and assessing the change in the measured noise power on a spectrum analyzer. This approach is commonly known as the Y-factor method and is described in more detail in the Methods. An acoustic noise figure of 2.8 ± 0.5 dB is achieved at 1.01 GHz with an electronic gain of 28 dB. We determined the standard deviation for the acoustic noise figure by assessing error propagation in our measurement as described in Supplementary Note 4. The corresponding terminal noise figure is 13.75 dB and is limited by the IDT losses.

Comparison of our measured noise figure to an acoustoelectric noise model based on diffusive processes[36] indicates that our minimum noise figure is limited by joule heating and the resulting thermal instabilities, as evidenced by the sudden increase in noise figure at high drift field, as shown in Fig. 3(b), that also corresponds to the drift field at which the gain becomes unstable. There are several paths forward to mitigate this effect, as will be discussed in the following section; however, one simple approach is to reduce the semiconductor conductivity $\sigma$ by lowering the amplifier layer doping concentration, which will lower the dissipated power required to achieve terminal gain. We anticipate that if thermal effects are mitigated, then the next dominating source of noise will be diffusion effects from semiconductor traps. As mentioned above, a theoretical treatment for the noise figure for an acoustoelectric amplifier including the impact of semiconductor traps has already been developed[36] based on a normal mode theory[35] and is discussed in Supplementary Note 5. This theoretical model suggests that it is possible to reduce our acoustic noise figure in this heterostructure to below 1 dB.

Lower IDT loss would also substantially reduce our terminal noise figure and should be achievable.[37-39] The IDTs in this work were far from ideal, being bidirectional and having a non-ideal duty cycle on account of being close to the resolution limit of the optical lithography system used to pattern them. However, in high $k^2$ systems such as this one, it is possible to optimize the insertion loss of an IDT to less than 1 dB.[37-39] Regardless, the noise figure we report here is the smallest acoustic and terminal noise figure ever achieved for an acoustoelectric amplifier and, with optimization of materials and IDTs, these devices could achieve a total noise figure less than 3 dB, making them competitive for integration into RF electronic systems.

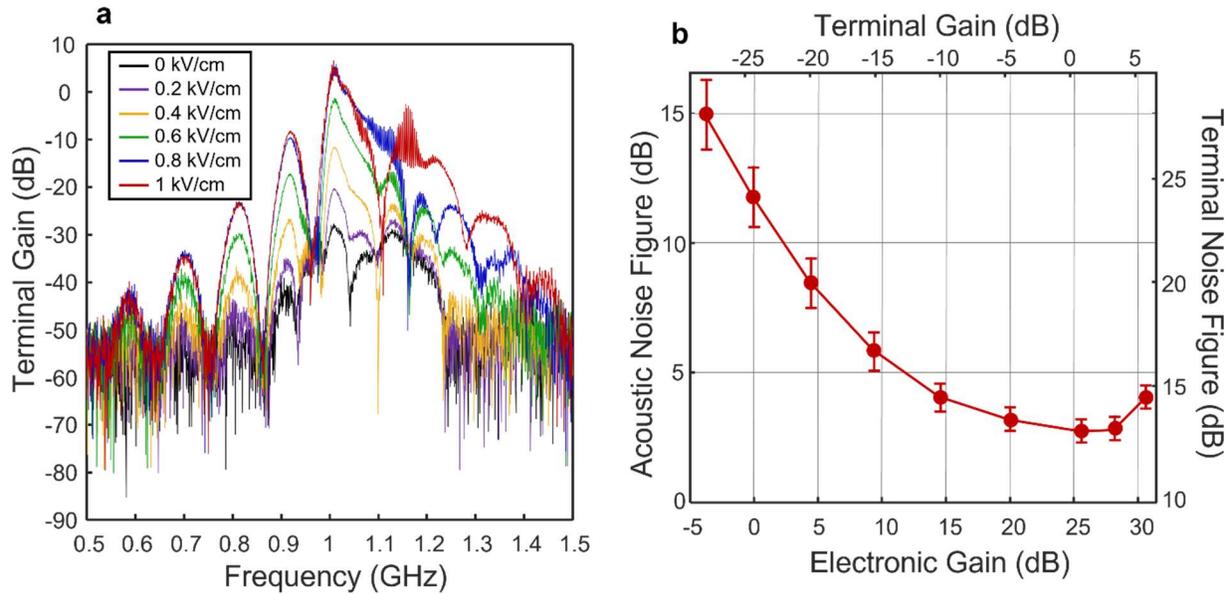

**Fig. 3: Noise figure measurement.** (a) Terminal gain as a function of frequency for increasing drift field values. A terminal gain of 6.5 dB is achieved at 1.01 GHz. For this device, the length is 500 μm, the acoustic wavelength is 3.75 μm, the IDT aperture is 56.25 μm, and each IDT has 10 electrode pairs. The measured noise figure is shown in (b) as a function of electronic and terminal gain. We demonstrate an acoustic noise figure of 2.8 ± 0.5 dB at 28 dB of electronic gain in this device. The error bars correspond to ± one standard deviation for the acoustic noise figure, which is determined via propagation of measurement error as described in Supplementary Note 4.

## Heterostructure design and optimization

In this section, we discuss the design and optimization of our acoustoelectric heterostructure in more detail. While here we focus on the specific $In_{0.53}Ga_{0.47}As$-$LiNbO_3$-silicon stack, the general three-layer heterostructure shown in Fig. 1(a) can lead to high-performance acoustoelectric devices for many choices of materials, as long as a similar approach is taken to optimize the parameters in each of the layers as well as the couplings between them.

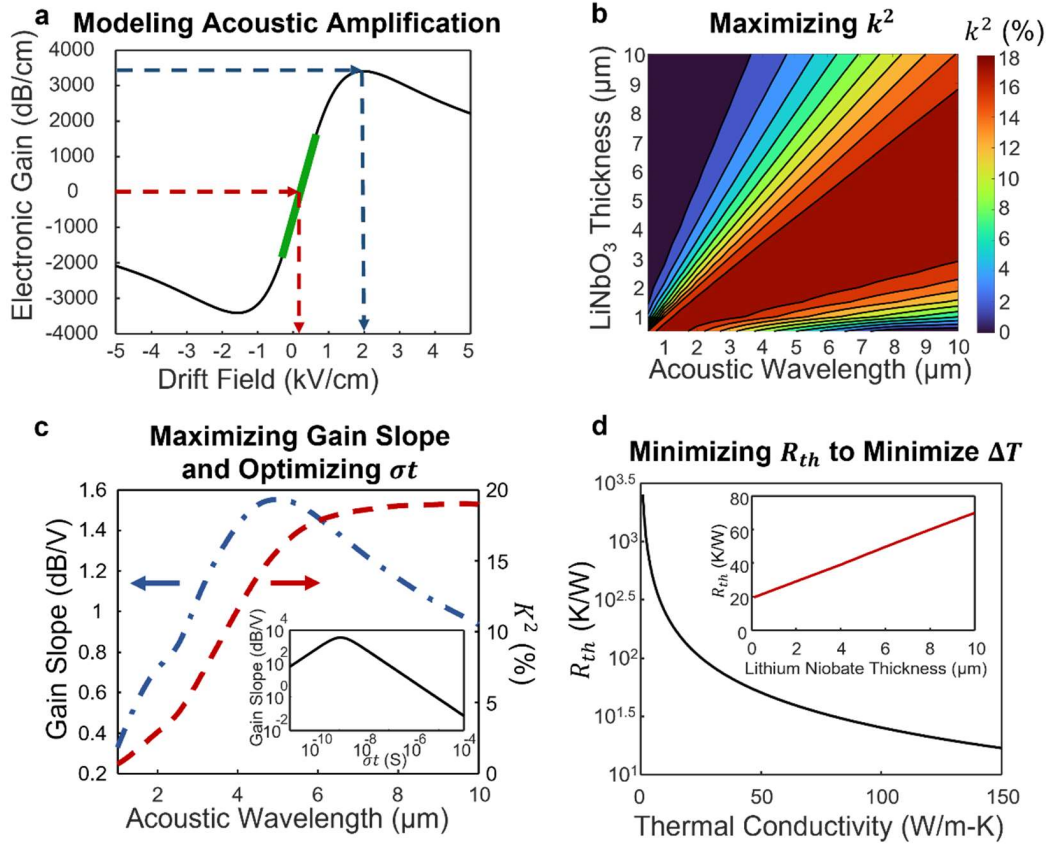

**Fig. 4: Heterostructure optimization.** (a) A plot of the theoretical electronic gain as a function of drift field for an equivalent circuit model of the acoustoelectric interaction. The red dashed line indicates the synchronous point where $v_d = v_a$, the dashed blue line indicates the maximum gain value, and the green line indicates the maximum gain slope. (b) Contour plot of the modeled $k^2$ value as a function of $LiNbO_3$ film thickness and acoustic wavelength. (c) The theoretical maximum gain slope and the modeled $k^2$ value are plotted as a function of acoustic wavelength for a $LiNbO_3$ film thickness of 5 μm. The inset shows a plot of the theoretical gain slope as a function of $\sigma t$. (d) A plot of the theoretical $R_{th}$ as a function of the substrate thermal conductivity for a heat source with a thickness of 50 nm, length of 500 μm, and width of 50 μm. The inset shows the computed $R_{th}$ for a heat transfer FEM model of the $In_{0.53}Ga_{0.47}As$- $LiNbO_3$-silicon heterostructure as a function of the $LiNbO_3$ thickness. In the model, the $In_{0.53}Ga_{0.47}As$ layer serves as the heat source with a thickness of 50 nm, length of 500 μm, and width of 50 μm. For a delay line amplifier built in the $In_{0.53}Ga_{0.47}As$-$LiNbO_3$-silicon heterostructure, a plot of the theoretical electronic gain as a function of drift field is shown in Fig. 4(a). Details on modeling of the electronic gain and a complete discussion of all material and device parameters are given in Supplementary Note 6. The maximum gain is set by $k^2$, but in practice the ability to achieve

this maximum is hindered by thermal effects. We then seek to optimize the maximum gain *slope* (gain per volt), which occurs around the drift field where the drift velocity ($v_d$) is equal to the acoustic wave phase velocity ($v_a$), to achieve large gain with minimal power dissipation. This requires maximizing $k^2$ while optimizing $\sigma t$ (in this case corresponding to minimizing $\sigma t$). As thermal effects have been an outstanding issue for the successful operation of active acoustic wave devices,[8-11] any practical design of these devices must also seek to minimize the temperature rise from power dissipation ($\Delta T$).

Maximizing $k^2$ is achieved first and foremost with the choice of piezoelectric material, propagation direction and polarization of acoustic waves in that material, and the frequency of operation relative to the thickness. Here we choose an SH-like guided mode propagating in the X material direction of Y-cut LiNbO$_3$ with particle motion primarily in the Z material direction, as this is analogous to YX SH-SAW modes in bulk LiNbO$_3$ that achieve the highest $k^2$ for bulk material but have high propagation losses due to coupling to bulk modes.[11] As the bulk modes are forbidden in this vertical guided-wave system, we use this mode to find a simultaneously low-loss and high-$k^2$ mode. Figure 4(b) shows a contour plot extracted from a FEM model of $k^2$ for the guided acoustic mode with primarily SH polarization in a film of YX LiNbO$_3$ on silicon, as a function of the LiNbO$_3$ thickness and acoustic wavelength. A large $k^2$ can be achieved for a significant range of acoustic wavelengths and LiNbO$_3$ thicknesses. At high operating frequencies, which correspond to small acoustic wavelengths, a high $k^2$ (>5%) can be achieved in this heterostructure with a 5 μm thick LiNbO$_3$ film by transitioning to a Rayleigh or SH-SAW quasi-surface mode as opposed to the guided acoustic mode used in this work. This can be used, for example, to achieve large, broadband gain using a single choice of film thickness.

Figure 4(c) shows the theoretical maximum gain slope and the modeled $k^2$ value as a function of acoustic wavelength for a fixed thickness of 5 μm, which is a common commercially available thickness for LiNbO$_3$-silicon wafers. The inset of Fig. 4(c) shows that the theoretical gain slope can reach exceptionally high values (>3500 dB/V) by modifying $\sigma t$. The discrepancy between the optimal gain slope and the gain slope we demonstrate here is due to challenges that limit semiconductor mobility, doping concentration, and thickness, as described in Supplementary Note 6. However, not all uses of acoustoelectric amplifiers favor the absolute maximum gain slope; for example, the saturation output power for an acoustoelectric amplifier is fundamentally limited by the capture of all available charge carriers in the acoustic wave potential and therefore is increased with a higher $N_d$,[35] showing that the optimization of the materials is actually application dependent.

As discussed above, it is critical to minimize the temperature rise from power dissipation, $\Delta T$, which is generally minimized by decreasing the heterostructure thermal resistance ($R_{th}$) and dissipated power. Figure 4(d) shows a plot of $R_{th}$ as a function of the substrate thermal conductivity and the inset of Fig. 4(d) shows a plot of $R_{th}$ as a function of the LiNbO$_3$ film thickness. Details on the thermal modeling are given in Supplementary Note 7. With 5 μm of LiNbO$_3$, we achieve a ~10X improvement in $R_{th}$ when compared to bulk LiNbO$_3$ and this can be further improved another 3x by reducing the LiNbO$_3$ thickness to 1 μm. Increasing the film resistivity or reducing the amplifier width (i.e., the width of the semiconductor amplifier region) also leads to a lower $\Delta T$ via reduced power dissipation that does not compromise gain (see Supplementary Note 7). However, these cannot necessarily be optimized independently in every case, as increasing the semiconductor resistivity is limited by the material platform and a narrow acoustic wave device can result in significant diffraction losses; for this reason, it is universally desirable to use a substrate with a high thermal conductivity for long delay lines.

**Conclusions**
In this article, we have presented an $In_{0.53}Ga_{0.47}As$-$LiNbO_3$-silicon heterostructure that enables the demonstration of a monolithic acoustoelectric amplifier operating continuously while achieving terminal gain. We demonstrate 11.25 dB of terminal gain at gigahertz frequencies in a 500 μm long amplifying delay line operating at 1 GHz, with <20 mW of dissipated power. We have also measured the acoustic noise figure of our platform to be 2.8 dB, which is the lowest noise figure ever reported for an acoustoelectric amplifier.

Together with our experimental results, we have presented modeling that suggest that there is potential for significantly higher performance; i.e., these devices can achieve higher gain in a shorter length, lower dissipated power, and lower noise figure. Thinning the $LiNbO_3$ layer will reduce $R_{th}$ and increase thermal stability without compromising gain. Decreasing the semiconductor conductivity, the thickness, or both will also lead to reduced power dissipation, improved thermal properties, and an increased gain slope. For example, simply reducing the doping concentration to $N_d = 5 \times 10^{15}$ cm$^{-3}$ should increase the demonstrated gain slope by >3X. Further reduction in the semiconductor amplifier width, especially when combined with acoustic guiding[40-43] or focusing[44-47] methods, could also significantly reduce the dissipated power and improve the overall device thermal stability. These improvements should also improve the noise figure as our acoustic noise figure is currently limited by the onset of thermal instability. Our terminal noise figure is currently limited by input losses and will improve with development of a low-loss IDT for $LiNbO_3$ films on silicon.

The heterostructure approach presented here can easily accommodate other material stacks. For example, lithium tantalate could be used as the piezoelectric film, which could potentially provide increased temperature stability.[48] Another option is aluminum nitride doped with scandium ($Sc_xAl_{1-x}N$); this is a deposited film, making it easier to achieve very thin films as well as easier and cheaper to manufacture. Moreover, $Sc_xAl_{1-x}N$ has a large acoustic velocity, which makes operating at gigahertz frequencies less constrained by IDT patterning capabilities.[49,50] A substrate, such as silicon carbide, could provide an even higher thermal conductivity than silicon and significantly improve the DC and RF power handling. The continued development of piezoelectric thin films, wafer bonding, and low defectivity material deposition and growth should continue to expand the possible material combinations to achieve high-performing active acoustic wave devices.

This work enables a new class of acoustoelectric devices at gigahertz frequencies with large gain, small footprint, large RF power handling, and low acoustic noise figure. While acoustic nonreciprocity can be achieved using acoustic nonlinearities,[21,22,51] the acoustic Zeeman effect,[52] nonreciprocal bianisotropy,[53,54] or spatiotemporal modulation of acoustic resonators,[55,56] here we show that exceptionally large nonreciprocal transmission (>44 dB) can be achieved at gigahertz frequencies using the acoustoelectric effect. Moreover, the nonreciprocity that we demonstrate is tunable, broadband, only requires an applied DC bias, and occurs in an on-chip device with a small footprint. Piezoelectric acoustic wave devices also allow coupling between electrical circuits and optomechanical devices, and recent experimental demonstrations have shown the potential advantage of implementing additional functionalities in the acoustic domain for classical and quantum information processing.[23,28,57] The acoustoelectric effect, using a heterostructure approach as described in this work, could then enable tunable acoustic loss (or gain) and tunable nonreciprocity in these hybrid electrical and piezo-optomechanical circuits. The acoustoelectric effect has been directly proposed for novel quantum information processing architectures using piezoelectric semiconductors.[26,30,31] Our heterostructure approach to acoustoelectric devices could enable unprecedented performance and new applications in these quantum systems. There is also potential to achieve net round-trip gain in a resonator using the material platform presented in this work, which could lead to on-chip frequency-selective resonant amplifiers and oscillators.[58] Together with other passive acoustic, active (nonlinear), and nonreciprocal acoustoelectric components,[10,17] these devices provide a library for integration in the front-end of RF electronic systems and miniaturized RF systems that utilize all-acoustic RF signal processing.

**Methods**
**Device fabrication.** A lattice-matched epitaxial semiconductor $In_{0.53}Ga_{0.47}As$/InP layered structure is first grown by MOCVD on a 2-inch InP substrate. The structure consists of a 500 nm NID InP buffer, a 3 μm NID $In_{0.53}Ga_{0.47}As$ etch stop, a 100 nm InP etch stop doped with silicon at $N_d = 1 \times 10^{18}$ cm$^{-3}$, a two-layer epitaxial contact (100 nm thick $In_{0.53}Ga_{0.47}As$ and 30 nm thick InP contact layers doped with silicon at $N_d = 2 \times 10^{19}$ cm$^{-3}$ and $N_d = 1 \times 10^{18}$ cm$^{-3}$, respectively), an $In_{0.53}Ga_{0.47}As$ amplifier layer with a target silicon doping level of $N_d = 1 \times 10^{16}$ cm$^{-3}$ and a target thickness of 50 nm, and a final 5 nm thick NID InP adhesion layer. The InP substrate is then bonded, at the wafer scale, to a 4-inch substrate consisting of a 5 μm thick LiNbO$_3$ film on bulk silicon. The bonding occurs via manual initiation followed by annealing at 100°C in vacuum. The InP substrate and buffer layer are then etched away in a hydrochloric acid (HCl) solution followed by removal of the $In_{0.53}Ga_{0.47}As$ etch stop layer in a solution of sulfuric acid (H$_2$SO$_4$), hydrogen peroxide (H$_2$O$_2$), and water (H$_2$O). The InP etch stop layer is then removed in a solution of HCl and phosphoric acid (H$_3$PO$_4$). The epitaxial contact is then patterned, and the wafer is placed in a H$_2$SO$_4$, H$_2$O$_2$, and H$_2$O solution to first etch the $In_{0.53}Ga_{0.47}As$ layer followed by a HCl and H$_3$PO$_4$ solution to etch the InP contact layer. Following patterning and etching of the epitaxial contact, the $In_{0.53}Ga_{0.47}As$ amplifier layer is patterned and etched in a H$_2$SO$_4$, H$_2$O$_2$, and H$_2$O solution, landing on the LiNbO$_3$. The IDTs are fabricated using a metal liftoff process where the pattern is made by photolithography or e-beam lithography depending on the required resolution and pattern quality. The IDT metal is 150 nm aluminum with a 10 nm chrome adhesion layer. A second liftoff step of a stack of 10 nm titanium, 500 nm gold, 500 nm silver, and 100 nm gold forms the metal contact. In the structure, the metal contact is fabricated so that it is not in the acoustic wave's path, which allows us to apply a drift field for the charge carriers while minimizing acoustic reflection loss.

**Passive delay line and acoustic gain measurements.** The delay line and acoustic amplifier devices are evaluated on a custom RF-DC probe station with separate ground-signal-ground (GSG) RF probes and DC probes. Scattering (S)-parameters are measured using a Keysight E5071c network analyzer. A two-port short-load-open-through (SLOT) calibration is performed using an impedance standard substrate before device measurements. To evaluate the acoustic amplifier performance, a continuous drift field is applied using a DC power supply while the S-parameters are measured as a function of frequency. The current is simultaneously measured with a source meter. The dissipated power is evaluated as the applied bias multiplied by the measured current. Terminal gain is the measured calibrated $S_{21}$ value on the network analyzer, meaning that for terminal gain to be achieved the acoustic wave amplification must overcome all device losses including the input/output losses at the IDTs. Electronic gain on a log scale ($G_{AE,dB}$) is given by $G_{AE,dB} = IL_{OFF} - IL_{ON} - \alpha_{AE}$ where $IL_{OFF}$ is the measured insertion loss on the network analyzer with no charge carrier drift field applied, $IL_{ON}$ is the measured insertion loss with the applied drift field, and $\alpha_{AE}$ is the loss due to the acoustoelectric effect with no applied drift field. The value for $\alpha_{AE}$ is determined by the drift field required to achieve the synchronous velocity condition $v_d = v_a$, which is specified by the measured Hall mobility and device length. In contrast to terminal gain, which only occurs when the gain is large enough to overcome end-to-end device losses, electronic gain is achieved when the drift velocity is large enough to overcome the synchronous condition of $v_d = v_a$.

**Determination of semiconductor and electromechanical properties.** The In$_{0.53}$Ga$_{0.47}$As semiconducting film is fully characterized by its thickness, mobility, and carrier concentration. The thickness of the patterned In$_{0.53}$Ga$_{0.47}$As amplifier layer was measured by profilometer after device fabrication at three different places on the wafer and the thickness was determined to be $58 \pm 3$ nm. The resistivity and Hall coefficient of the In$_{0.53}$Ga$_{0.47}$As amplifier layer was measured by the Van der Pauw method using Hall structures patterned on the wafer such that they go through the exact same fabrication process flow as the amplifier devices. Measurements were made on two separate Hall structures using a Bio-Rad fixed magnetic field Hall effect measurement system. The average values for the resistivity and Hall coefficient were $0.048 \pm 0.001$ Ω-cm and $-(3.5 \pm 0.1) \times 10^3$ m²/C, respectively. The measured Hall mobility was $4220 \pm 40$ cm²/V-s and the charge carrier concentration was $(3.1 \pm 0.1) \times 10^{16}$ cm³. The experimental data near the synchronous point was fit to the theoretical gain slope with $k^2$ as a fitting parameter to determine $k^2 = 14 \pm 1\%$.

**Noise figure measurement.** The noise figure was determined by measuring the noise temperature of the acoustic wave amplifier by the Y-factor method on a Rohde and Schwarz FPL1007 spectrum analyzer with an internal preamp using a calibrated noise source with an excess noise ratio (ENR) of 26 dB. The spectrum analyzer is designed specifically to measure noise figure and can perform a second-order calibration to remove the effect of the noise temperature for the spectrum analyzer, itself, which we have done in all measurements presented here. We then directly measured the terminal noise temperature for our acoustic wave amplifier, which includes the impact of the IDT input and output to the acoustoelectric interaction region. The noise factor $F$ can be calculated from the noise temperature of a system $T_{sys}$ according to

$F = (T_0 + T_{sys})/T_0$ where $T_0$ is the input noise temperature. We refer our noise to a room temperature of $T_0 = 290$ K. The noise figure $NF$ is $F$ expressed in dB (instead of on a linear scale), such that $NF = 10\log_{10}(F) = 10\log_{10}(1 + T_{sys}/290\text{ K})$. The Y-factor method determines the noise temperature and gain of a system by measuring the linear noise power on a spectrum analyzer for a low temperature noise source ($T_{source}^{OFF}$), which corresponds to when the calibrated noise source is off, and a high temperature noise source ($T_{source}^{ON}$), which corresponds to when the calibrated noise source is on. The Y-factor term, $Y$, is given by the ratio of the linear noise power with the noise source on ($N_{on}$) and off ($N_{off}$) such that $Y = N_{on}/N_{off}$ and the noise temperature is given by $T_{sys} = \frac{T_{source}^{ON} - Y \cdot T_{source}^{OFF}}{Y-1}$. To ensure device stability, each measurement was taken three times and the reported values are the average from the three measurements. From these measured values, we can then determine the acoustic noise figure by considering the losses present in our system. As measured on the spectrum analyzer and confirmed by measurement on a network analyzer, our total loss with no applied drift field is 28 dB. The loss due to the acoustoelectric effect with no applied drift field is 4 dB and the acoustic propagation loss, including the impact of the semiconductor film, is 2 dB. The combined loss from the input and output is then 22 dB. We split the remaining losses in half and assign one half to the input losses and the other half to the output losses. This approach was confirmed by (1) comparing the $S_{11}$ and $S_{22}$ values on the network analyzer and (2) measuring the noise figure with the input and output ports switched. To evaluate the acoustic noise figure, which corresponds to the noise figure of the acoustoelectric interaction region, we use Friis's formula for the noise temperature of a system $T_{sys} = T_1 + \frac{T_2}{G_1} + \frac{T_3}{G_1 G_2} = T_{IDT,in} + \frac{T_{AE}}{G_{IDT,in}} + \frac{T_{IDT,out}}{G_{IDT,in} G_{AE}}$ where $T_1 = T_{IDT,in}$ and $G_1 = G_{IDT,in}$ are the noise temperature and linear power gain, respectively, associated with the lossy input IDT, $T_2 = T_{AE}$ and $G_2 = G_{AE}$ are the noise temperature and linear power gain, respectively, associated with the acoustoelectric amplification region, and $T_3 = T_{IDT,out}$ is the noise temperature of the lossy output IDT.[59] The acoustic noise figure is then $NF_{acoustic} = 10\log_{10}(1 + T_{AE}/290\text{ K})$.

**Acknowledgements**
Supported by the Laboratory Directed Research and Development program at Sandia National Laboratories, a multimission laboratory managed and operated by National Technology and Engineering Solutions of Sandia LLC, a wholly owned subsidiary of Honeywell International Inc. for the U.S. Department of Energy's National Nuclear Security Administration under contract DE-NA0003525. This work was performed, in part, at the Center for Integrated Nanotechnologies, an Office of Science User Facility operated for the U.S. Department of Energy (DOE) Office of Science. This paper describes objective technical results and analysis. Any subjective views or opinions that might be expressed in the paper do not necessarily represent the views of the U.S. Department of Energy or the United States Government.

**Conflict of interest**
The authors declare that they have no conflict of interest.

# Supplementary Information

**Supplementary Note 1**
**Gain Compression**

Supplementary Fig. 1(a) shows a plot of the measured transmission as a function of frequency and applied bias for an acoustoelectric amplifier with a length of 250 µm, interdigital transducer (IDT) aperture of 240 µm, and acoustic wavelength of 16 µm. Terminal gain is not achieved in this device, but there is a significant increase in transmission around the resonance frequency of 250 MHz with an increasing applied bias. A plot of the electronic gain at 250 MHz as a function of the applied drift field is shown in Supplementary Fig. 1(b) for the forward propagating ($S_{21}$) and backward propagating ($S_{12}$) acoustic waves. Nonreciprocal amplification is observed as expected with a transmission contrast of 23.5 dB at a drift field of 1.8 kV/cm.

A plot of the acoustic output power as a function of the acoustic input power for various applied drift field values is shown in Supplementary Fig. 1(c). The acoustic input power is determined by subtracting the IDT input loss from the radio frequency input power set on the network analyzer. The acoustic output power is the measured electronic gain added to the acoustic input power. As can be seen in Supplementary Fig. 1(c), there is a gain compression of 1 dB at an acoustic output power of 12 dBm and drift field of 1.8 kV/cm.

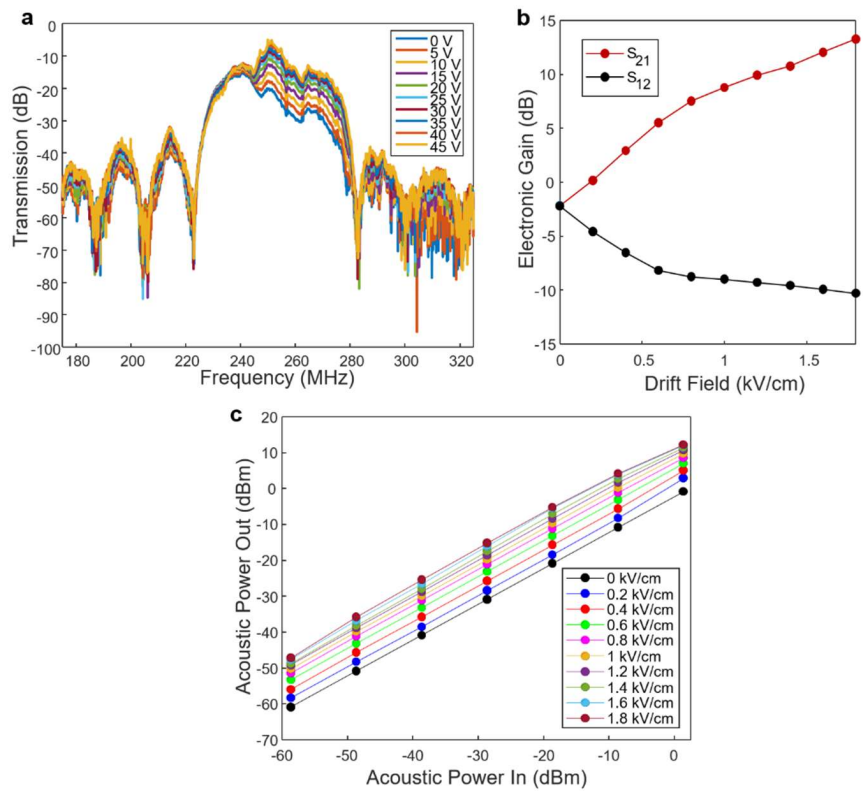

**Supplementary Figure 1. Gain compression.** (a) Measured transmission as a function of frequency for increasing applied bias values for an acoustoelectric amplifier with an acoustic wavelength of 16 µm. (b) Electronic gain for the forward propagating ($S_{21}$) and backward propagating ($S_{12}$) acoustic waves as a function of the applied drift field. (c) Acoustic output power as a function of acoustic input power for different applied drift fields.

## Supplementary Note 2
## Adhesion Layer Thickness

In our heterostructure, the indium gallium arsenide ($In_{0.53}Ga_{0.47}As$) semiconductor and lithium niobate ($LiNbO_3$) piezoelectric films are spaced by a non-intentionally doped (NID) indium phosphide (InP) adhesion layer that is 5 nm thick. Supplementary Fig. 2 shows a plot of the theoretical maximum electronic gain as a function of the operating frequency for various InP adhesion layer thicknesses. For this plot, we assume that the electromechanical coupling coefficient ($k^2$) is constant at 1% across all frequencies to isolate the impact of the adhesion layer thickness alone. As can be seen, a thicker adhesion layer reduces the achievable maximum electronic gain at high frequencies. The onset of gain reduction corresponds to where the ratio between the adhesion layer thickness and the acoustic wavelength becomes large (>0.01). Given that the electronic gain generally increases with frequency, as shown in Supplementary Fig. 2, the adhesion layer thicknesses is a controllable parameter to potentially achieve an acoustoelectric amplification that is constant with respect to frequency over a wide bandwidth.

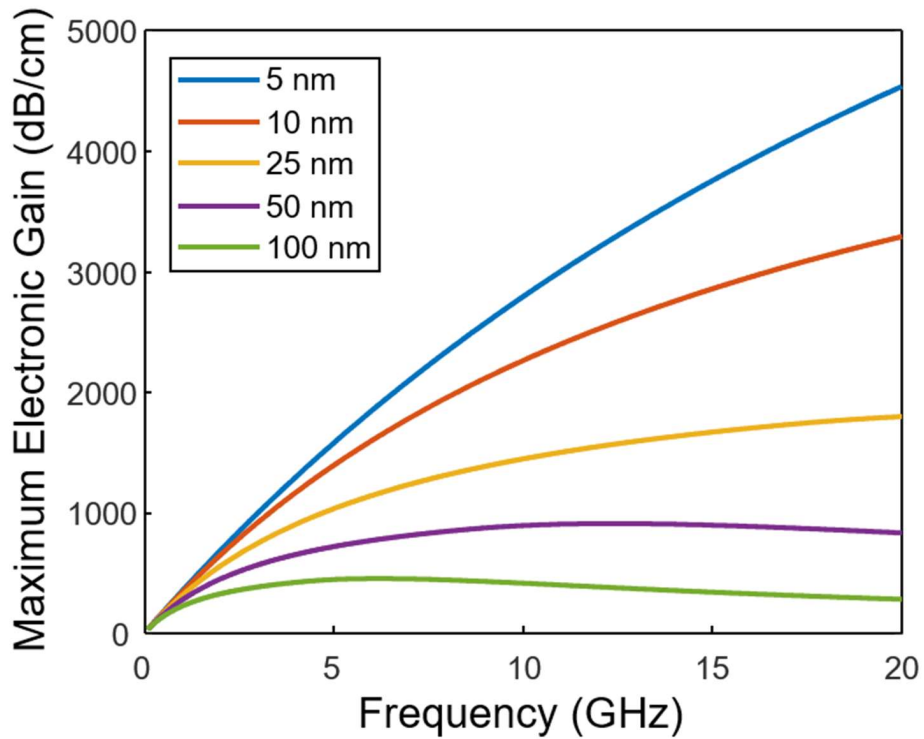

**Supplementary Figure 2. Adhesion layer thickness.** The theoretical maximum gain value is plotted as a function of the acoustic frequency for different adhesion layer thicknesses assuming a constant $k^2$ of 1%.

## Supplementary Note 3
## Experimental Nonreciprocity

Supplementary Fig. 3 shows the measured $S_{21}$ and $S_{12}$ values at 1.01 GHz as a function of the drift field for the charge carriers. A terminal gain of 6.5 dB is achieved at a drift field of 1 kV/cm and the transmission difference exceeds 44 dB between the forward and backward propagating acoustic waves, therefore showing that the gain is nonreciprocal. The experimentally achieved isolation for the backward propagating acoustic wave is limited due to the presence of additional acoustic modes with significantly lower $k^2$. The acoustic modes with low $k^2$ will only be weakly amplified or attenuated with an applied bias and lead to a constant background. Additional optimization of the LiNbO$_3$ thickness can likely reduce the effect of these modes. The acoustic transmission was not measured past a drift field of 1 kV/cm for this device as the S$_{21}$ plateau, shown in Supplementary Fig. 3, indicates an onset of thermal effects that can cause measurement instabilities and ultimately device damage. The onset of thermal effects ultimately prohibits operating at the maximum gain and further mitigation of these effects could enable significantly larger gain.

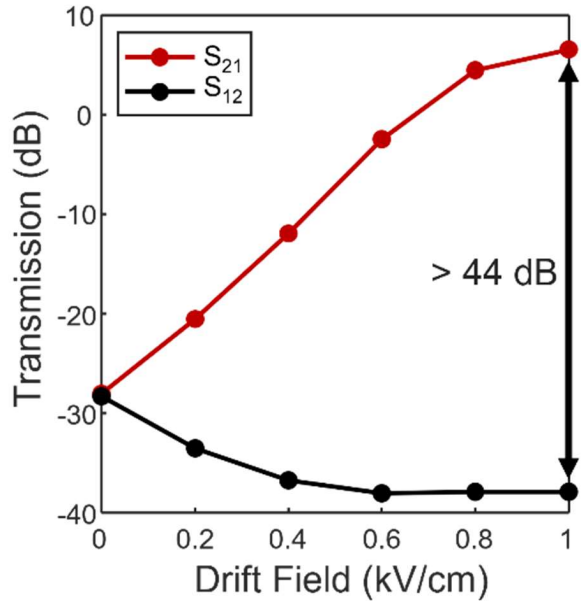

**Supplementary Figure 3. Nonreciprocal gain.** The measured S$_{21}$ and S$_{12}$ values at 1.01 GHz are plotted as a function of drift field for an amplifying delay line with a length of 500 μm, acoustic wavelength of 3.75 μm, and IDT aperture of 56.25 μm. A transmission difference exceeding 44 dB is achieved at a drift field of 1 kV/cm.

**Supplementary Note 4**
**Acoustic Noise Figure Error Propagation**
The expression for the acoustic noise factor ($F_{\text{acoustic}}$) is given by
$$F_{\text{acoustic}} = 1 + {T_{AE}}/{290 \text{ K}} \tag{S1}$$
while the acoustic noise figure ($NF_{\text{acoustic}}$) is given by
$$NF_{\text{acoustic}} = 10\log_{10}\left(1 + {T_{AE}}/{290 \text{ K}}\right) \tag{S2}$$
where $T_{AE}$ is the noise temperature of the acoustoelectric interaction region. The expression for $T_{AE}$ is
$$T_{AE} = T_{\text{sys}} G_{\text{IDT,in}} - T_{\text{IDT,in}} G_{\text{IDT,in}} - \frac{T_{\text{IDT,out}}}{G_{AE}} \tag{S3}$$

where $T_{sys}$ is the total measured noise temperature, $G_{IDT,in}$ is the linear power gain for the input IDT, $T_{IDT,in}$ is the noise temperature for the input IDT, $T_{IDT,out}$ is the noise temperature for the output IDT, and $G_{AE}$ is the linear electronic gain. The input and output IDTs to the acoustoelectric interaction region are lossy such that $G_{IDT,in}$ is less than one. To determine the error in $NF_{acoustic}$, we must then determine the error in $T_{AE}$. The variance of $T_{AE}$ $(\sigma_{T_{AE}})^2$ is

$$(\sigma_{T_{AE}})^2 = (\sigma_{T_{sys}})^2 \left(\frac{\partial T_{AE}}{\partial T_{sys}}\right)^2 + (\sigma_{G_{IDT,in}})^2 \left(\frac{\partial T_{AE}}{\partial G_{IDT,in}}\right)^2 + (\sigma_{T_{IDT,in}})^2 \left(\frac{\partial T_{AE}}{\partial T_{IDT,in}}\right)^2 \quad (S4)$$
$$+ (\sigma_{T_{IDT,out}})^2 \left(\frac{\partial T_{AE}}{\partial T_{IDT,out}}\right)^2 + (\sigma_{G_{AE}})^2 \left(\frac{\partial T_{AE}}{\partial G_{AE}}\right)^2$$

where $\sigma_{T_{sys}}$ is the standard deviation of $T_{sys}$, $\sigma_{G_{IDT,in}}$ is the standard deviation of $G_{IDT,in}$, $\sigma_{T_{IDT,in}}$ is the standard deviation of $T_{IDT,in}$, $\sigma_{T_{IDT,out}}$ is the standard deviation of $T_{IDT,out}$, and $\sigma_{G_{AE}}$ is the standard deviation of $G_{AE}$. Plugging in for the partial derivatives, we have

$$(\sigma_{T_{AE}})^2 = (\sigma_{T_{sys}})^2 G_{IDT,in}^2 + (\sigma_{G_{IDT,in}})^2 (T_{sys} - T_{IDT,in})^2 + (\sigma_{T_{IDT,in}})^2 G_{IDT,in}^2 + \quad (S5)$$
$$(\sigma_{T_{IDT,out}})^2 \left(\frac{1}{G_{AE}}\right)^2 + (\sigma_{G_{AE}})^2 \left(\frac{T_{IDT,out}}{G_{AE}^2}\right)^2.$$

We now consider the uncertainty of each variable. The value for $\sigma_{T_{sys}}$ is the standard deviation from three consecutive measurements for $T_{sys}$ taken on the spectrum analyzer. The uncertainty varies and is less than 1% when $T_{sys}$ reaches its minimum value. The value for $G_{IDT,in,dB}$, which is $G_{IDT,in}$ expressed in dB, is found according to $G_{IDT,in,dB} = -\frac{1}{s}(IL_{OFF} - \alpha_{AE} - \alpha_{prop})$ where $s$ is the splitting ratio of the loss between the input and output IDT, $IL_{OFF}$ is the insertion loss with no drift field applied, $\alpha_{AE}$ is the loss due to the acoustoelectric effect with no drift field applied, and $\alpha_{prop}$ is the propagation loss. As the input and output IDTs are identical, we assume that $s = 2$. However, fabrication imperfections can cause the input and output IDT to have different insertion loss values. The variance of $G_{IDT,in,dB}$ $\left((\sigma_{G_{IDT,in,dB}})^2\right)$ is given by

$$(\sigma_{G_{IDT,in,dB}})^2 = (\sigma_{IL_{OFF}})^2 \left(\frac{1}{s}\right)^2 + (\sigma_{\alpha_{AE}})^2 \left(\frac{1}{s}\right)^2 + (\sigma_{\alpha_{prop}})^2 \left(\frac{1}{s}\right)^2 + (\sigma_s)^2 \left(\frac{IL_{OFF}}{s^2} + \frac{\alpha_{AE}}{s^2} + \frac{\alpha_{prop}}{s^2}\right)^2$$

where $\sigma_{IL_{OFF}}$ is the standard deviation for $IL_{OFF}$, $\sigma_{\alpha_{AE}}$ is the standard deviation for $\alpha_{AE}$, and $\sigma_{\alpha_{prop}}$ is the standard deviation for $\alpha_{prop}$. The measurement of $IL_{OFF}$ is done on a network analyzer with an error that is at least one order of magnitude less than other errors present such that we can consider $\sigma_{IL_{OFF}} \approx 0$. The value for $\alpha_{AE}$ is found according to the experimental applied bias ($V$) required to achieve the synchronous condition of $v_d = v_a$ where $v_d$ is the charge carrier drift velocity and $v_a$ is the acoustic phase velocity. This applied bias is given by $V = v_a l/\mu$ where $l$ is the device length and $\mu$ is the semiconductor mobility. The degree of uncertainty in $V$ then corresponds to the degree of uncertainty in $\alpha_{AE}$. The value for $v_a$ is determined by the experimental measurement of the IDT resonance frequency and the acoustic wavelength, which is defined by the IDT pitch. The IDT resonance frequency is measured on a network analyzer to a high precision (<<1% of uncertainty) and the IDT pitch uncertainty is determined by the patterning uncertainty of the photolithography system, which is <<1%. In contrast, the experimental uncertainty in the measured mobility is 1% and we calculate the uncertainty in $\alpha_{AE}$ to be 1%. The value for $\sigma_{\alpha_{\text{prop}}}$ is 1 dB, found by the uncertainty in the linear fitting while the value for $\sigma_s$ is found according to the difference in the $S_{11}$ and $S_{22}$ values measured on the network analyzer on resonance. From this, we then determine the standard deviation for $G_{\text{IDT,in}}$ ($\sigma_{G_{\text{IDT,in}}}$). $T_{\text{IDT,in}}$ and $T_{\text{IDT,out}}$ are determined according to $G_{\text{IDT,in}}$ and therefore the degree of uncertainty in $G_{\text{IDT,in}}$ determines $\sigma_{T_{\text{IDT,in}}}$ and $\sigma_{T_{\text{IDT,out}}}$. The variance for the electronic gain in dB $\left(\sigma_{G_{AE,\text{dB}}}\right)^2$ is $\left(\sigma_{G_{AE,\text{dB}}}\right)^2 = \left(\sigma_{IL_{OFF}}\right)^2 + \left(\sigma_{IL_{ON}}\right)^2 + \left(\sigma_{\alpha_{AE}}\right)^2$ given that $G_{AE,\text{dB}} = IL_{OFF} - IL_{ON} - \alpha_{AE}$. As $IL_{OFF}$ and $IL_{ON}$ are both highly accurate measurements made on a network analyzer, we can consider $\left(\sigma_{IL_{OFF}}\right)^2 \approx \left(\sigma_{IL_{ON}}\right)^2 \approx 0$. The uncertainty in the electronic gain is then determined by the uncertainty in the loss due to the acoustoelectric effect with no drift field applied, which is 1% as described above. It then follows that the error in $\sigma_{G_{AE}}$ is approximately 1%.

From the uncertainty of each value, we then calculate $\left(\sigma_{T_{AE}}\right)^2$ according to Eqn. (S5). From Eqn. (S1) we have the following for the variance in $F_{\text{acoustic}}$ $\left(\left(\sigma_{F_{\text{acoustic}}}\right)^2\right)$

$$\left(\sigma_{F_{\text{acoustic}}}\right)^2 = \left(\sigma_{T_{AE}}\right)^2 \left(\frac{1}{290 \text{ K}}\right)^2. \tag{S6}$$

To convert the error in $F_{\text{acoustic}}$ to a standard deviation for $NF_{\text{acoustic}}$ ($\sigma_{NF_{\text{acoustic}}}$) we use the following:

$$\sigma_{NF_{\text{acoustic}}} = 10\log_{10}(1 + F_{\text{acoustic,error}}) \tag{S7}$$

where $F_{\text{acoustic,error}}$ is the percentage error in $F_{\text{acoustic}}$ calculated according to $\sigma_{F_{\text{acoustic}}}/F_{\text{acoustic}}$ where $\sigma_{F_{\text{acoustic}}}$ is the standard deviation of $F_{\text{acoustic}}$ calculated from Eqn. (S6). Currently, the uncertainty in the minimum value for $NF_{\text{acoustic}}$ is primarily limited by the uncertainty in the splitting ratio between the input and output IDTs and the uncertainty in the propagation loss measurement.

**Supplementary Note 5**
**Theoretical Acoustic Noise Figure**
A theoretical analysis of the noise figure for acoustic wave amplifiers based on the acoustoelectric effect already exists and has been successfully used to fit experimental data for a separated-medium amplifier,[36] which uses isolated piezoelectric and semiconductor materials separated by an air gap. The noise figure calculation uses the impedance field method and a normal mode formula for describing the interaction between the charge carrier system and the acoustic wave. The expression for $F_{\text{acoustic}}$ is given by

$$F_{\text{acoustic}} = 1 + \frac{\exp(-2k_a h)}{(v_d/v_a - 1)} \left(\frac{1 + \varepsilon_p/\varepsilon_h \tanh(k_a h)}{1 + \tanh(k_a h)}\right)^2 \left(1 + \frac{D_{TR}}{D_{TH}}\right) \left(\frac{\exp(2\alpha l) - 1}{\exp(2\alpha l)}\right) \tag{S8}$$

where $k_a$ is the acoustic propagation constant, $h$ is the distance separating the semiconductor and piezoelectric materials, $\varepsilon_p$ is the piezoelectric permittivity, $\varepsilon_h$ is the permittivity of the material between the semiconductor and piezoelectric, $D_{TR}$ is a diffusion constant to account for semiconductor trapping effects, $D_{TH}$ is the thermal diffusion constant, $\alpha$ is the electronic gain, calculated according to normal mode theory,[35] and $l$ is the device length. The trapping constant is calculated according to $D_{TR} = \frac{f(\omega)[1-f(\omega)]v_d^2 \tau_c}{1+(\omega\tau_c)^2}$ where $f(\omega)$ is the fraction of total charge carriers that are untrapped, $\omega$ is the carrier wave frequency, $\tau_c$ is the relaxation time for carrier trapping, and $D_{TH} = \frac{k_B T \mu}{q}$ where $k_B$ is Boltzmann's constant, $T$ is the temperature, and $q$ is the elementary charge.

The theoretical acoustic noise figure for our heterostructure, according to Eqn. S8, is plotted in Supplementary Fig. 4 as a function of electronic gain for several different values for $f(\omega)$. For this plot, $h = 5$ nm, $v_a = 3750$ m/s, $\sigma t = 100$ μS, $k^2 = 10\%$, $\omega/2\pi = 1$ GHz, $\tau_c \omega = 1$, and $T = 290$ K. Both the value of the acoustic noise figure and the dependence of the noise figure on the electronic gain varies with $f(\omega)$. The increase in the noise figure at large electronic gain, seen in Supplementary Fig. 4 for all values of $f(\omega)$ except for $f(\omega) = 1$, occurs due to the increasing dominance of the $D_{TR}/D_{TH}$ term at larger drift velocities. As can be seen in Supplementary Fig. 4, theoretically the acoustic noise figure can be less than 1 dB in this heterostructure and it is advantageous to achieve $f(\omega) \sim 1$ where trapping effects are negligible.

Over a wide range of fitting parameters, this model for the acoustic noise figure does not fit our experimental data. We find that our experimental data deviates from both the expected gain model and noise figure model due to heating effects. Despite this deviation, we find this model of the noise figure to be useful to assess what the noise figure could be if thermal effects are mitigated and the noise is dominated by diffusive effects.

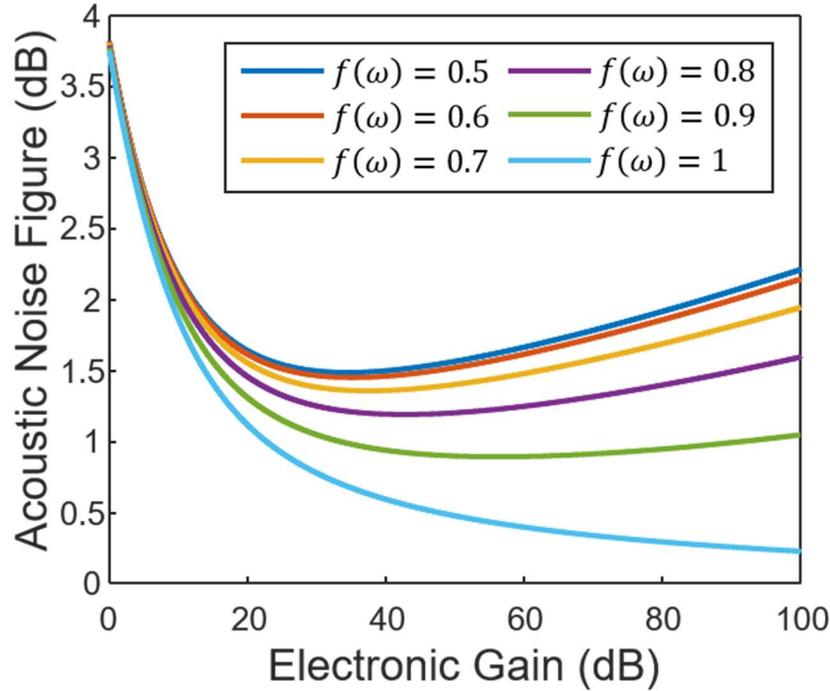

**Supplementary Figure 4. Noise figure model.** A plot of the theoretical acoustic noise figure for our heterostructure is shown as a function of electronic gain for different values of $f(\omega)$.

**Supplementary Note 6**
**Gain Curve Modeling and Critical Material Parameters for High Performance**
Electronic gain as a function of the applied carrier drift field, under the assumption of small-signal operation with no thermal effects, can be well-described by a model that treats the acoustoelectric interaction as an equivalent RC circuit.[60] From this model, the gain $\alpha$ is given by

$$\alpha = -\frac{1}{2}k^2 k_a \frac{\gamma\omega\tau}{1+(\gamma\omega\tau)^2} \quad (S9)$$

where $\gamma = 1 - v_a/v_d$, $\omega$ is the acoustic frequency, and $\tau$ is the effective RC time constant (dielectric relaxation time) given by $\tau = \frac{\varepsilon_p + \varepsilon_0}{\sigma t k_a}$ where $\varepsilon_0$ is the vacuum permittivity.[60] In this equivalent circuit model, a gap between the piezoelectric and the semiconductor materials (such as the bonding adhesion layer in our case) provides a series capacitance that can be corrected by dividing $\tau$ and $k^2$ by a frequency-dependent reduction factor $1 + k_a h\, \varepsilon_p/\varepsilon_h$.[60]

There are several important aspects of the theoretical gain curve to consider. The first is that with no drift field the charge carriers lag the acoustic wave, which results in acoustic attenuation. A nonzero drift field must be applied to achieve the synchronous condition where the carriers drift at a velocity equal to the acoustic phase velocity. Only when the carriers drift at a velocity exceeding the acoustic phase velocity is electronic gain achieved in the system. The required drift field ($E_d$) to achieve the conditions where $v_d = v_a$ is given by $E_d = v_a/\mu$. The semiconducting material should have a high mobility such that a lower voltage is required to achieve gain, both because high-voltage operation is generally disfavored and more practically because ohmic heating typically limits the performance of such devices, as it does for the devices studied in this work. Simultaneously, the conductivity-thickness product of the semiconductor film ($\sigma t$), as well as all material dielectric constants, determine an effective RC time constant ($\tau$) that produces a frequency-dependent acoustoelectric response and can optimize the acoustoelectric interaction for a given frequency. The defectivity of the semiconductor should be minimized to enable a low noise figure, as defectivity leads to an electronic diffusivity that induces broadband amplitude and phase noise.[36] The interaction strength between the charge carriers and the acoustic wave is specified by a coupling coefficient ($k^2$) that characterizes the ratio of stored mechanical energy to input electrical energy when the system is electrically driven.[61] The piezoelectric material thus needs to support an acoustic mode with a high $k^2$ and, ideally, low acoustic loss to minimize the degree of amplification needed to overcome system losses. The substrate material should have a high thermal conductivity to minimize temperature rise for a given power dissipation ($\Delta T$) and therefore enable active acoustic wave devices to operate continuously and stably. In addition, the acoustic velocity in the substrate must be larger than in the piezoelectric material, as this supports a guided acoustic wave in the piezoelectric material that suppresses acoustic radiation (loss) into the substrate. Finally, because the piezoelectric acoustic wave's electric field overlaps the substrate, the substrate must be a low loss electrical material at the operating radio frequency (RF) (small loss tangent). In this supplementary note, we discuss the semiconducting and piezoelectric properties for our material stack. The following supplementary note discusses thermal modeling and the reduction of $\Delta T$.

Supplementary Fig. 5 shows a contour plot of the dissipated DC power required to achieve 30 dB of electronic gain as a function of the semiconductor $\sigma t$ and μ. As can be seen, the dissipated DC power is minimized by simultaenously achieving a low $\sigma t$ and high μ. In the actual metal organic chemical vapor deposition (MOCVD) growth of the lattice-matched indium gallium arsenide/indium phosphide (In$_{0.53}$Ga$_{0.47}$As/InP) structure, it is challenging to separately optimize $\sigma t$ and µ in this way.[62] However, we can see that µ should ideally at least exceed 1000 cm$^2$/V-s. A reasonable thickness for the In$_{0.53}$Ga$_{0.47}$As layer is approximately 50 nm to minimize surface scattering and a reasonable doping concentration is 1x10$^{16}$ cm$^{-3}$. Within these constraints, the expected gain slope for our heterostructure is approximately 2 dB/V.

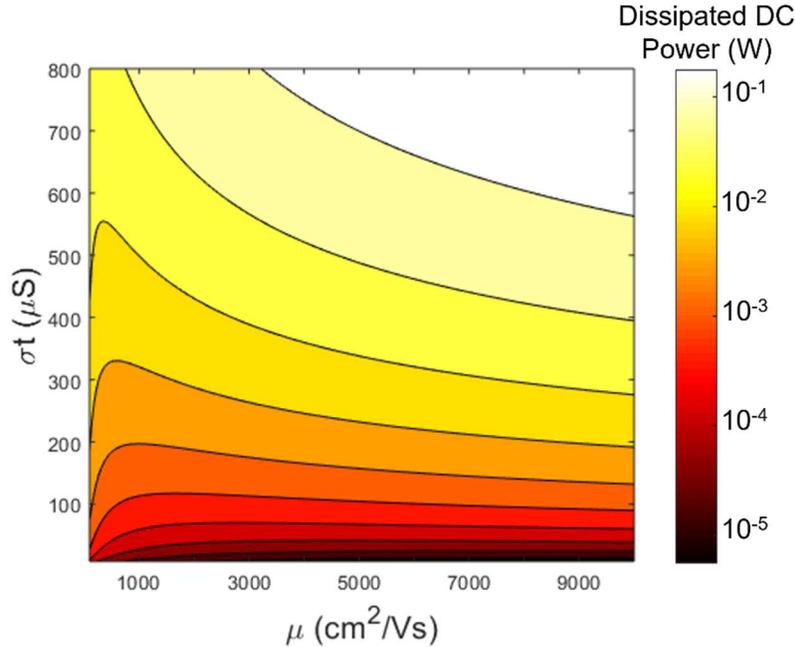

**Supplementary Figure 5. Mobility optimization.** A contour plot of the dissipated power required to achieve 30 dB of electronic gain as a function of $\sigma t$ and µ.

LiNbO$_3$ is a piezoelectric material that is well-known to support acoustic modes with high $k^2$ and low loss both as a bulk material and a suspended thin film.[10,11,37,48] Non-suspended LiNbO$_3$ films on silicon is a relatively unexplored material platform for piezoelectric acoustic wave devices. In this material stack, acoustic modes are guided in the LiNbO$_3$ due to the slower speed of sound in LiNbO$_3$ compared to silicon. The measured S$_{11}$ and S$_{21}$ as a function of frequency for a fabricated delay line device with an acoustic wavelength of 3.75 µm on a YX LiNbO$_3$ film on bulk silicon is shown in Supplementary Fig. 6(a). There are two clear resonances; one occurs at approximately 1 GHz and corresponds to an acoustic mode with primarily shear horizontal (SH) polarization and high $k^2$ (>10%), which results in the strongly coupled mode. The other occurs at approximately 3.4 GHz and corresponds to the third-harmonic of this fundamental mode. Other acoustic modes are expected to propagate in the delay line, but have low $k^2$ (<1%) and are only weakly coupled.

There are no reported measurements of acoustic propagation loss at gigahertz frequencies for a LiNbO$_3$ film on silicon and the expected acoustic losses are difficult to model. Therefore, we experimentally assess the acoustic losses by fabricating delay line devices with varying gap lengths. The measured insertion loss as a function of gap length is shown in Supplementary Fig.

6(b). The data at 1.075 GHz and 3.4 GHz was taken from the fundamental and third harmonic resonances on the same devices with an acoustic wavelength of 3.75 µm. The data at 0.77 GHz was taken from a different set of delay lines with an acoustic wavelength of 5 µm. For all delay line devices, the IDT had 10 electrode pairs and an aperture of 100 µm. From Supplementary Fig. 6(b), we extract a propagation loss of 35 dB/cm at 1.075 GHz, 68 dB/cm at 3.35 GHz, and 117 dB/cm at 0.77 GHz. The primary contribution to the propagation loss is likely scattering effects in the film and at the LiNbO$_3$-silicon interface, which depend on the operating frequency and acoustic mode profile within the material stack.

In the main article, we report the modeling of $k^2$ for this guided mode with primarily SH polarization. This coupling coefficient was found via finite element method (FEM) simulation by calculating the change in acoustic velocity that occurs between the case of free propagation and the case of a perfectly conducting boundary condition on the surface. We find that $k^2$ exceeds 10% over a large range of acoustic wavelengths. We report our measurement of the acoustic propagation loss in the LiNbO$_3$-silicon platform to be 35 dB/cm at approximately 1 GHz; since, as we show, large gain can be achieved with amplifier lengths substantially less than 0.1 cm, these films are effectively both low loss in addition to supporting acoustic modes with high $k^2$.

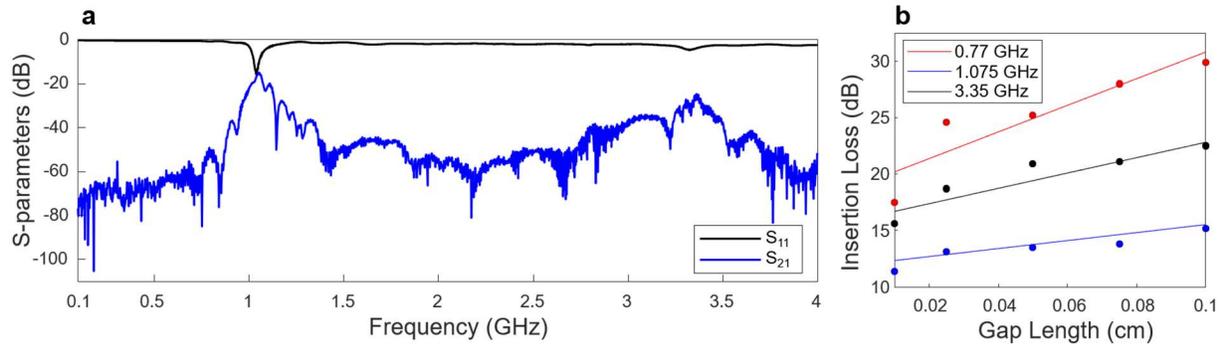

**Supplementary Figure 6. Assessment of acoustic losses in a LiNbO$_3$ film on bulk silicon piezoelectric substrate.** (a) The measured S$_{11}$ and S$_{21}$ as a function of frequency for a fabricated delay line on the YX LiNbO$_3$ film on bulk silicon. (b) A plot of insertion loss as a function of gap length for delay line devices with varying lengths.

**Supplementary Note 7**
**Thermal Modeling**
The thermal resistance ($R_{th}$) can be calculated by an analytical model[63] or computed by a heat transfer FEM model, which allows more complex device geometry, including modeling the effects of a multilayered substrate. The FEM model consists of a 500 µm x 50 µm x 50 nm In$_{0.53}$Ga$_{0.47}$As heat source, on a LiNbO$_3$ film of variable thickness on a 500 µm thick silicon substrate. For calculating $R_{th}$ values, the size of the modeled substrate is 50 cm x 50 cm to eliminate any edge effects. The model includes both the effects of thermal conduction and convection with a heat transfer coefficient of 5 W/m$^2$-K. The In$_{0.53}$Ga$_{0.47}$As, LiNbO$_3$, and silicon have thermal conductivity values of 5 W/m-K, 4.6 W/m-K, and 148 W/m-K, respectively.

We find that for a substrate that is entirely LiNbO3, we expect $R_{th}$ = 553 K/W, while we expect $R_{th}$ = 17 K/W for an entirely silicon substrate. In the case of a film of LiNbO3 on silicon, the $R_{th}$ value varies with the LiNbO3 film thickness. With 5 μm of LiNbO3, the modeled $R_{th}$ value is 44 K/W, meaning that with the silicon substrate we achieve a ~10X improvement in $R_{th}$ when compared to bulk LiNbO3.

Increasing the film resistivity or reducing the semiconductor width leads to reduced dissipated power and therefore a reduced $\Delta T$. Supplementary Fig. 7(a) shows a log-log plot of the modeled $\Delta T$ as a function of $\sigma t$. The FEM model consists of a 500 μm x 50 μm x 50 nm In$_{0.53}$Ga$_{0.47}$As heat source on a 5 μm thick LiNbO3 film on a 500 μm thick silicon substrate. The size of the modeled substrate is 2.5 x 2.5 cm. It is advantageous to minimize $\sigma t$ to minimize $\Delta T$, yet we must work within the material platform limitations on $\sigma t$ as discussed in Supplementary Note 6. In this heterostructure, the $\Delta T$ also decreases as a function of the semiconductor width due to reduced dissipated power, as shown in Supplementary Fig. 7(b). However, a narrow acoustic wave device can result in significant diffraction losses and therefore there is an optimization required to reduce the semiconductor width while avoiding excessive acoustic losses.

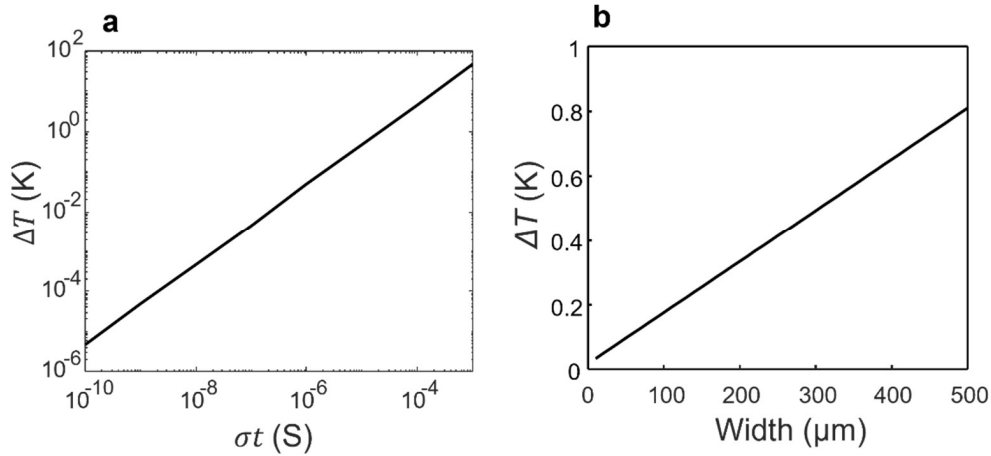

**Supplementary Figure 7. Thermal modeling.** Plots of $\Delta T$ as a function of (a) $\sigma t$ and (b) semiconductor width.